\documentclass[useAMS]{mn2e}
\usepackage{psfig}

\usepackage{times}
\usepackage{amssymb,amsmath}

\title[Axions from the Sun?]{No axions from the Sun}

\author[Roncadelli \& Tavecchio]
{
M. Roncadelli$^1$ and F. Tavecchio$^2$\\
$^1$INFN -- Sezione di Pavia, Via A. Bassi 6, I -- 27100 Pavia, Italy\\
$^2$INAF -- Osservatorio Astronomico di Brera, via E. Bianchi 46, I--23807 Merate, Italy\\
}

\begin{document}

\maketitle

\begin{abstract}
Preliminary evidence of solar axions in XMM-{\it Newton} observations has quite recently been claimed by Fraser et al. as an interpretation of their detection of a seasonally-modulated excess of the X-ray background. Within such an interpretation, these authors also estimate the axion mass to be $m_a \simeq 2.3 \cdot 10^{- 6} \, {\rm eV}$. Since an axion with this mass behaves as a cold dark matter particle, according to the proposed interpretation the considered detection directly concerns cold dark matter as well. So, the suggested interpretation would lead to a revolutionary discovery if confirmed. 
Unfortunately, we have identified three distinct problems in this interpretation of the observed result of Fraser et al. which ultimately imply that the detected signal -- while extremely interesting in itself -- cannot have any relation with hypothetical axions produced by the Sun. Thus, a physically consistent interpretation of the observed seasonally-modulated X-ray excess still remains an exciting challenge.
\end{abstract}


\section{Introduction}

In spite of the fact that cold dark matter is a compelling requirement of the standard cosmological model, no convincing evidence for its detection exists to date. Among so many candidates for cold dark matter particles, the neutralino and the axion are certainly the most popular ones (for a review, see: Bertone 2010). We stress that an axionic cold dark matter would also naturally solve the long-standing strong CP problem (for a review, see: Kim 1987, Cheng 1988, Kim \& Carosi 2010).

Quite recently, Fraser et al. published the observational evidence of a seasonally-modulated X-ray background in excess of the cosmic one, obtained after a careful analysis of the XMM-{\it Newton} data accumulated in the period 2000-2012 (Fraser et al. 2014). Having convincingly demonstrated the existence of this background, they proposed that it originates from the conversion in the Earth magnetosphere of axions produced in the Sun. Their claim seems to be supported by the nice agreement between the spectrum of the detected X-ray component with that expected from the conversion of axions produced in the solar core through Primakoff effect, Compton effect and electron Bremsstrahlung effect (Redondo 2013).

Given the importance of these topics, a very careful scrutiny of the interpretation put forward by Fraser et al. looks compelling. Accomplishing this task is precisely the aim of the present Letter. We are ultimately led to the conclusion that it is {\it impossible}  that the signal detected by the XMM-{\it Newton} observatory is related to axions potentially emitted from the Sun and thus with cold dark matter. 

\section{Discussion of the axion interpretation}

In order to facilitate the reader's understanding, we start by schematically summarizing the main argument of Fraser et al. (2014).

Actually, the {\it observational result} -- against which we have no objection whatsoever -- is the evidence of a seasonally modulated background in excess of the instrumental background and the constant cosmic X-ray background (CXB). The authors make a big effort to show that this effect is not an artifact of the detectors but a physical result. 

The proposed {\it interpretation} of this observational evidence -- which is the focus of our work -- can be broadly sketched through the following chain of arguments.

\begin{itemize} 

\item The Sun is supposed to isotropically emit axions owing to three different production channels (among others which are irrelevant for the present discussion): Primakoff effect, Compton effect and electron Bremsstrahlung effect (Redondo 2013). Below roughly $4.8 \, {\rm keV}$ electron Bremsstrahlung dominates, and it is with this part of the spectrum that Fraser et al. 
are mainly concerned. Because axion production in the Sun occurs in the core, it can be regarded for all practical purposes as a point-like axion source.

\item According to a suggestion previously put forward by Davoudiasl and Huber (2006, 2008), these axions are supposed to convert to X-rays in the geomagnetic field. Note that these authors supposed as usual that an X-ray is collinear with the parent axion.

\item The XMM-{\it Newton} observatory obviously never points towards the Sun, in order to avoid immediate destruction of its 
X-ray detectors. Instead, during its orbit it points in a varying direction always quite different from the line of sight to the Sun. So, the obvious question arises: how is it possible that the X-rays originating from the conversion of solar axions in the geomagnetic field can enter the XMM detectors all the time?

\item Fraser et al. answer such a question by invoking a result of Guendelman and collaborators (Guendelman et al. 2008, 2010, 2012. See also Redondo 2010). Basically, the latter authors show that, when an axion-to-photon conversion takes place in an inhomogeneous magnetic field, then the produced photon can be non-collinear with the parent axion. To be sure, this result has been proved only for very special configurations of the magnetic field and what happens in the geomagnetic one is totally unknown. As these authors themselves remark, this is just the analog of the Stern-Gerlach effect for photons or axions (depending on which particle is produced). 

\item For the sake of illustration, we find it convenient to deal with a much more familiar situation: we consider the {\it Gedanken} experiment in which solar axions are replaced by electrons and the XMM detectors by charged-particle detectors. That is to say, we imagine that the Sun emits electrons rather than axions isotropically from the core and with the same flux as that assumed for axions (of course, we neglect interactions among electrons). In such a situation, the geomagnetic field -- being extremely complicated -- certainly possesses a gradient, so that the usual Stern-Gerlach effect takes place. Therefore, the electrons from the Sun will be effectively isotropized, thereby giving rise to a background wherein electrons move along any direction. Note that the flux of such a background is {\it strongly reduced} with respect to the solar flux in the absence of the Stern-Gerlach effect~\footnote{This circumstance is analogous to the fact that in the usual Stern-Gerlach setup concerning the splitting of a beam of spin $s$ atoms into $2s+1$ beams, the intensity of each split beam is $2s+1$ times smaller than the intensity of the original beam.}. Moreover, XMM has a very small field of view, which entails that only a {\it very small fraction} of the electron background can be detected. Returning now to the real case of axions, Fraser et al. explicitly state: ``It is thought here that isotropic scattering axion-to-photon conversion probabilities can attain values of the same order as for purely collinear scattering''. We denote by $\Pi$ the {\it isotropization parameter}, which quantifies the photon flux reduction arising from the ``photonic'' Stern-Gerlach effect, and by $\xi$ the {\it geometric factor} that accounts for the very small field of view of XMM. As usual, $\xi$ is defined as
\begin{equation} 
\xi \equiv \frac{\Omega_{\rm XMM}}{\Omega_{\rm SC}}~, \ \ \ \ \ \ \ \ \ \ \ \ \ \ \Omega_{\rm XMM} < \Omega_{\rm SC}~,
\label{a1}
\end{equation} 
where $\Omega_{\rm XMM}$ is the aperture of XMM and $\Omega_{\rm SC}$ the scattering solid angle, namely the solid angle encompassing the total detectable flux (obviously we have $\xi = 1$ for $\Omega_{\rm XMM} \geq \Omega_{\rm SC}$). Manifestly, the result of Fraser et al. should depend on both $\Pi$ and $\xi$.

\item As we said, Fraser et al. focus their attention on the solar axion flux below $4.8 \, {\rm keV}$, where the electron Bremsstrahlung is the dominant process, which is obviously controlled by the axion-electron coupling constant $g_{ae}$. Nevertheless, they also take into account the sub-leading contributions from the Compton emission and the Primakoff processes, which depend on the axion-photon coupling constant $g_{a \gamma}$. In order to fit their observed spectrum they need to assume $g_{ae} \simeq 2.2 \cdot 10^{- 12}$ and $g_{a \gamma} \simeq 10^{-10} \, {\rm GeV}^{- 1}$. 

\item Fraser et al. also attempt to make an estimate of the axion mass. Because the region where they are most sensitive is around $10^{- 6} \, {\rm eV}$, they conclude that the axion mass has to be $m_a \simeq 2.3 \cdot 10^{- 6} \, {\rm eV}$. This is a value for which the axion is a very good candidate for cold dark matter (for a review, see: Bertone 2010). 

\end{itemize}

% ---------------------------------------------------
\begin{figure}
%\hskip -2 cm
%\hspace*{-0.3 truecm}
\vspace*{-1.6 truecm}
\hspace*{-0.4 truecm}
%\vskip -0.3 cm
\psfig{file=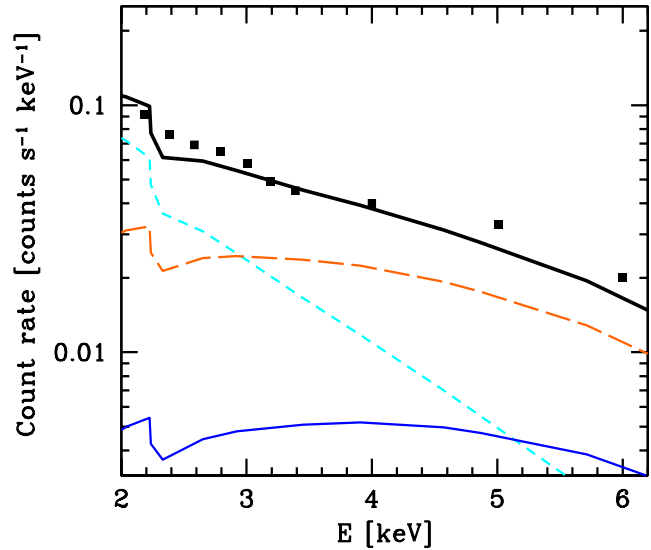,height=10.5cm,width=10.2cm}
\vspace{-1.3 cm}
\caption{EPIC pn difference spectrum (black points, from Fig. 20 of Fraser et al.) with the expected X-ray converted spectrum from solar axion (solid black line) calculated as detailed in the text with $g_{ae} \simeq 2.2 \cdot 10^{-12}$ and $g_{a \gamma} \simeq10^{-10} \, {\rm GeV}^{-1}$. The long-dashed orange, the dashed cyan and the solid blue lines show the contributions from the Primakoff, electron Bremsstrahlung and Compton emission processes, respectively.}
\label{fig:spec}
\end{figure}
% ---------------------------------------------------

\noindent Let us now outline our criticism. 

We start from what we regard as the main point. Fraser et al. based their interpretation on Fig. 20 of their paper, in which they show that the spectrum of the seasonally-modulated excess background can be reproduced by the expected spectrum of the solar axions. Because it is not clear where the parameters $\Pi$ and $\xi$ enter the calculations leading to Fig. 20, we have attempted to reproduce it, taking as a starting point $\Pi = \xi = 1$. Their observed spectrum of time-dependent excess background is taken from that Figure and reported in our Fig. 1 as black squares. Next, using their equations we have evaluated the flux contributions from Primakoff, electron Bremsstrahlung and Compton emission processes according to their choice of the relevant parameters, namely $g_{ae} \simeq 2.2 \cdot 10^{-12}$ and $g_{a \gamma} \simeq 10^{-10} \, {\rm GeV}^{- 1}$, and the axion-to-X-ray conversion probability in the geomagnetic field: they are shown in Fig. 1 by the long-dashed orange, the dashed cyan and the solid blue lines, respectively. Finally, we have summed them getting the solid black line in Fig. 1, which turns out to {\it exactly match} the data points and in fact coincides with the fitting line in Fig. 20 of Fraser et al. for the same values of the parameters. The crux of the argument is that we have exactly reproduced what  Fraser et al. provided that $\Pi = \xi = 1$. This is the proof that their interpretation hold true {\it only under the impossible assumption that XMM-Newton points directly towards the Sun and that the axion-to-photon conversions are fully collinear}. 

Moreover, we show that if the solar axion flux were fully isotropized by the invoked ``photonic'' Stern-Gerlach effect then XMM would detect no signal whatsoever. Indeed, in such a situation we would have $\Omega_{\rm SC} = 4 \pi$ and denoting by $\theta_{\rm XMM}$ the field of view of XMM we get $\Omega_{\rm XMM} = \pi \, \theta^2_{\rm XMM}$. Because $\theta_{\rm XMM} \simeq 30 \, {\rm arcmin}$, Eq.~(\ref{a1}) yields $\xi \simeq 10^{- 5}$. So, we do not even need to estimate the isotropization parameter $\Pi$ to prove our conclusion.

A second weak point of the Fraser et al. interpretation is the specific choice of the axion-electron coupling constant $g_{ae} \simeq 2.2 \cdot 10^{- 12}$, which exceeds by about a factor of $5$ the most recent upper bound $g_{ae} < 4.3 \cdot 10^{- 13}$ (Viaux et al. 2013). A similar problem -- even if less significant -- concerns the axion-photon coupling constant $g_{a \gamma} \simeq 10^{- 10}$, since the most recent upper bound is $g_{a \gamma} < 0.6 \cdot 10^{- 10}$ (Ayala et al. 2014).

A third critical point concerns the estimated value of the axion mass $m_a \simeq 2.3 \cdot 10^{- 6} \, {\rm eV}$. Actually, it is well known that in {\it any} axion model its mass $m_a$ is strictly related to the axion-photon coupling constant $g_{a \gamma}$ by the equation 
\begin{equation}
m_a = 0.7 \, \beta \left(\frac{g_{a \gamma}}{10^{- 10} \, {\rm GeV}^{- 1}} \right) {\rm eV}~, 
\label{a2} 
\end{equation}
where $\beta$ is a model-dependent constant of order 1 (Cheng et al. 1995). As a consequence, with the choice of Fraser et al.  the axion mass would be $m_a = 0.7 \, \beta \, {\rm eV} \sim 1 \, {\rm eV}$ and the axions would behave as {\it hot} dark matter (Turner 1987, 1988; Mass\'o et al. 2002; Graf \& Steffen 2011). So, if they were the dominant component of the dark matter, then the structure formation in the Universe would be impossible (for a review, see: Mo, van den Bosch \& White 2010).
 
In conclusion, we have demonstrated that the interpretation advocated by Fraser et al. of their observational result rises quite severe problems to make it untenable. Throughout this Letter, we have attempted our best to distinguish between {\it observations} and {\it interpretation}. As a consequence, the observed seasonally-modulated X-ray excess looks like a very important discovery. And a physically consistent interpretation thereof still remains an exciting challenge.

\section*{Acknowledgments}
We would like to thank the referee Konstantin Zioutas and the Editor 
for their advise which helped us quite a lot in the improvement of the 
presentation of our result. The work of M. R. is supported by the INFN grants TaSP and CTA.


\begin{thebibliography}{}

\bibitem[]{mirizzigamma} Ayala A. et al., arxiv:1406.6053.

\bibitem[]{bertone} Bertone G. (Ed.), 2010, {\it Particle Dark Matter} (Cambridge University Press, Cambridge, 2010).

\bibitem[]{} Cheng H. Y.,1988, Phys. Rep. 158, 1

\bibitem[]{cgn} Cheng S. L., Geng C. Q., Ni W. T. 1995, Phys. Rev. D 52, 3132

\bibitem[]{huber} Davoudiasl H. and Huber P., 2006, Phys. Rev. Lett. 97, 141302

\bibitem[]{huber2} Davoudiasl H. and Huber P., 2008, JCAP 08, 026

\bibitem[]{fraser} Fraser G.W., et al. 2014, MNRAS 445, 2146

\bibitem[]{ahem} Graf P., Steffen F. D. 2011, Phys. Rev. D 83, 075011

\bibitem[]{guendelman} Guendelman E. I. 2008, Phys. Lett. B 662, 445

\bibitem[]{guendelman}  Guendelman E. I., Shilon I., Cantatore G., and Zioutas, K. 2010, JCAP 6, 031
 
\bibitem[]{guendelman}  Guendelman E. I., Leizerovich E. I., Shilon I. 2012, Int. J. Mod. Phys. 27, 1252018

\bibitem[]{} Kim J. H., 1987, Phys. Rep. 150, 1

\bibitem[]{} Kim J.E., Carosi G., 2010,  Rev. Mod. Phys. 82, 557.

\bibitem[]{ahem} Mass\'o E., Rota F. and Zsembinszki G. 2002, Phys. Rev. D 66, 023004

\bibitem[]{ahem}  Mo H., van den Bosch F. and White S., 2010, {\it Galaxy Formation and Evolution}, (Cambridge University Press, Cambridge, 2010).

\bibitem[]{redondo} Redondo J., 2010, arxiv:1003.0410. 

\bibitem[]{redondo} Redondo J., 2013, , JCAP 12, 008

\bibitem[]{ahem} Turner M. S. 1987, Phys. Rev. Lett. 59, 2489

\bibitem[]{ahem} Turner M. S. 1988, Phys. Rev. Lett.  60, 1101 

\bibitem[]{viaux} Viaux N. et al., 2013, Phys. Rev. Lett. 111,  231301

\end{thebibliography}
\end{document}